\documentclass[a4paper,preprintnumbers,showpacs,twocolumn,superscriptaddress,nofootinbib,amsmath,amssymb,floatfix]{revtex4-1}

\usepackage{amsmath,amssymb,graphicx,amsfonts}
\usepackage{mathrsfs}
\usepackage{comment}
\usepackage{xcolor}
\usepackage[shortlabels]{enumitem}
\usepackage{url}

\usepackage{graphicx}

\usepackage{dcolumn}
\usepackage{bm}
\usepackage{wasysym}
\usepackage{color}
\usepackage{multirow}
\usepackage{booktabs}
\usepackage{float}

\begin{document}
\def\Dt{\Delta t}
\def\Rdot{\stackrel{\cdot }{\relR}}
\def\Rddot{\stackrel{\cdot \cdot }{\relR}}


\title{\textbf{Investigation of the formation of superheavy elements with atomic numbers 116 and 120 through $^{50}$Ti-induced reaction}}

\author{Bakhodir Kayumov}
\email{b.kayumov@newuu.uz} 
\affiliation{New Uzbekistan University, Tashkent 100000, Uzbekistan}
\affiliation{Institute of Nuclear Physics, Tashkent 100214, Uzbekistan}



\begin{abstract}
The synthesis of superheavy elements provides crucial insights into the stability and structure of nuclei at the limits of the periodic table. This study investigates the $^{50}$Ti$+^{244}$Pu and $^{50}$Ti$+^{251}$Cf reactions as pathways to form superheavy elements (SHEs) 116 (Livermorium) and 120, respectively. Using the dinuclear system model, key parameters such as fusion probability and fusion cross section were calculated. These findings were used to examine neutron emission and survival probability through a statistical approach. The reaction $^{50}$Ti$+^{244}$Pu is explored as a continuation of experimental efforts to extend the known isotopic range of element 116, while the $^{50}$Ti$+^{251}$Cf reaction represents a frontier for the synthesis of element 120. The role of shell effects, excitation energy, and angular momentum on the production and stability of these nuclei is discussed. The results provide theoretical predictions to guide future experimental efforts aimed at advancing our understanding of the island of stability and the limits of nuclear existence.
\end{abstract}

\maketitle
\section{Introduction}

The synthesis of superheavy elements (SHEs) is a crucial area of research in nuclear physics, providing insights into the stability and structure of nuclei at the extreme limits of the periodic table \cite{Hofman2000, Giuliani2019}. The production of SHEs is particularly challenging due to the interplay between fusion, fission, and quasifission processes, which determine the probability of forming a compound nucleus and its subsequent survival against fission. Recent experimental advancements by using $^{48}$Ca projectiles have contributed significantly to understanding the underlying mechanisms governing SHE formation \cite{Stavsetra2009, Oganessian2010, Dullmann2010, Oganessian2013, Oganessian2013_2, Khuyagbaatar2014, Utyonkov2015}. In pursuit of reaching island of stability, there have been a lot of theoretical studies aimed at predicting the production of 119 and 120 elements using actinide targets with the projectiles $^{50}$Ti, $^{51}$V, $^{54}$Cr and $^{58}$Fe \cite{Nasirov2011, Wang2011, Siwek2012, Kuzmina2012, Liu2013, Zhu2014, ZAGREBAEV2015, Adamian2020, Li2022120, Zhu2023, Khuyagbaatar2022, Zhang2024, Nasirov2024, Kayumov2022}. Author in \cite{Long022030} proposed analytical formula that could be universally applied in hot fusion reactions for calculating the optimal incident energies (OIE) and predicted the OIE for synthesizing 119 and 120 elements by using different projectile-target combination. The results of these studies can be used to find optimal entrance parameter such as projectile-target combination, optimal colliding energies for the production of SHE.

The present study investigates the formation of superheavy elements with atomic numbers 116 (Livermorium) and 120 through the $^{50}$Ti+$^{244}$Pu and $^{50}$Ti+$^{251}$Cf fusion reactions. Reactions with $^{50}$Ti serve as potential pathways for extending the known isotopic range of Livermorium and exploring the opportunity of synthesizing element 120 \cite{Nasirov2011, Li2018, Niu2021}. The choice of these target-projectile combinations is motivated by lack of the theoretical predictions and recent experimental studies at Lawrence Berkeley National Laboratory \cite{ Gates2024}.

To model the reaction dynamics, the dinuclear system (DNS) model employed, which provides a comprehensive framework for calculating the capture cross section, compound nucleus formation probability, and evaporation residue (ER) cross sections. This approach accounts for nuclear structure effects, including deformation and shell corrections, which significantly influence the fusion probability \cite{Adamian2020, Kayumov2022, Nasirov2024, Nasirov2024232TH}.

Furthermore, the study incorporates recent refinements in statistical models to evaluate neutron emission and survival probabilities, using the KEWPIE2 code \cite{Kewpie2}. By considering energy-dependent fission barrier modifications, shell corrections, and orientation effects, calculation has an aim to provide theoretical guidance for future experimental investigations \cite{Giardina2018, Nasirov2024}. The results presented in this work contribute to the ongoing search for new superheavy elements and enhance our understanding of the limits of nuclear stability. They may serve as a theoretical benchmark for upcoming experiments at leading research facilities, such as the RIKEN, GSI and other high-intensity heavy-ion accelerator laboratories worldwide.

On the other hands, orientation effects on fusion were investigated in heavy ion reactions using deformed actinide nuclei as target, by \cite{Nishio2000, Oganessian2004PRC, Nishio2012}. These experimental studies concluded that equatorial collisions makes significant contribution to the fusion whereas polar collisions do not lead to fusion. In this work, dependence of the fusion cross section on orientational angle was studied for $^{50}$Ti+$^{244}$Pu and $^{50}$Ti+$^{251}$Cf reactions. 

This paper is structured as follows: Section II discusses the capture cross section calculations, Section III presents the compound nucleus formation probabilities, and Section IV analyzes fusion and evaporation residue cross sections. Finally, the conclusions are outlined in Section VI.

\section{Capture cross section}
The capture of an incoming projectile-nucleus by the target-nucleus is a necessary condition in the formation of the fusion. Therefore, it is essential to calculate in which initial quantities the system goes to capture and not. To find values for the energy and angular momentum where the capture occurs, the equation for the relative motion of colliding nuclei can be used \cite{Giardina2000,Nasirov2005}. 
\begin{eqnarray}
 \label{maineq} &&\frac{d \dot R}{dt} +
 \gamma_{R}(R,\alpha_1,\alpha_2)\dot R(t)= F(R,\alpha_1,\alpha_2),\nonumber\\
 \label{maineq2} &&F(R,\alpha_1,\alpha_2)=
 -\frac {\partial V(R,\ell,\alpha_1,\alpha_2)}{\partial R}-
 \dot R^2 \frac {\partial \mu(R)}{\partial R}\,\nonumber\\
 \label{maineq3}&&\frac{dL}{dt}=\left(\dot{\theta}
 R(t) -\dot{\theta_1} R_{1eff} -\dot{\theta_2} R_{2eff}\right)\nonumber \\
 &&\hspace{0.625 cm}\times\gamma_{\theta}(R,\alpha_1,\alpha_2)R(t),\\
 &&L_0=J_R(R,\alpha_1,\alpha_2) \dot{\theta}+J_1 \dot{\theta_1}+J_2 \dot{\theta_2}\,\nonumber\\
 &&E_{rot}=\frac{J_R(R,\alpha_1,\alpha_2) \dot{\theta_{}}{}^2}2+\frac{J_1
 \dot{\theta_1}^2}2+\frac{J_2 \dot{\theta_2}^2}2\nonumber.
 \end{eqnarray}
In the equation above, the nucleus-nucleus potential $V(R,\ell,\alpha_1,\alpha_2)$ was calculated as a sum of Coulomb, nuclear, and rotational potentials \cite{Giardina2000,Nasirov2005}. The nuclear part of the nucleus-nucleus potential is calculated using the folding procedure between the effective nucleon-nucleon forces suggested by Migdal \cite{Migdal1983}. In estimation of nucleus-nucleus potential quadrupole and octupole deformation are used as deformation parameters $\beta_i$ of the colliding deformed nuclei $i=1$ and 2, and their values are taken from \cite{AUDI2003, MolNix1995} tables. 

\begin{figure}[ht]
\centering
\includegraphics[width=0.50\textwidth]{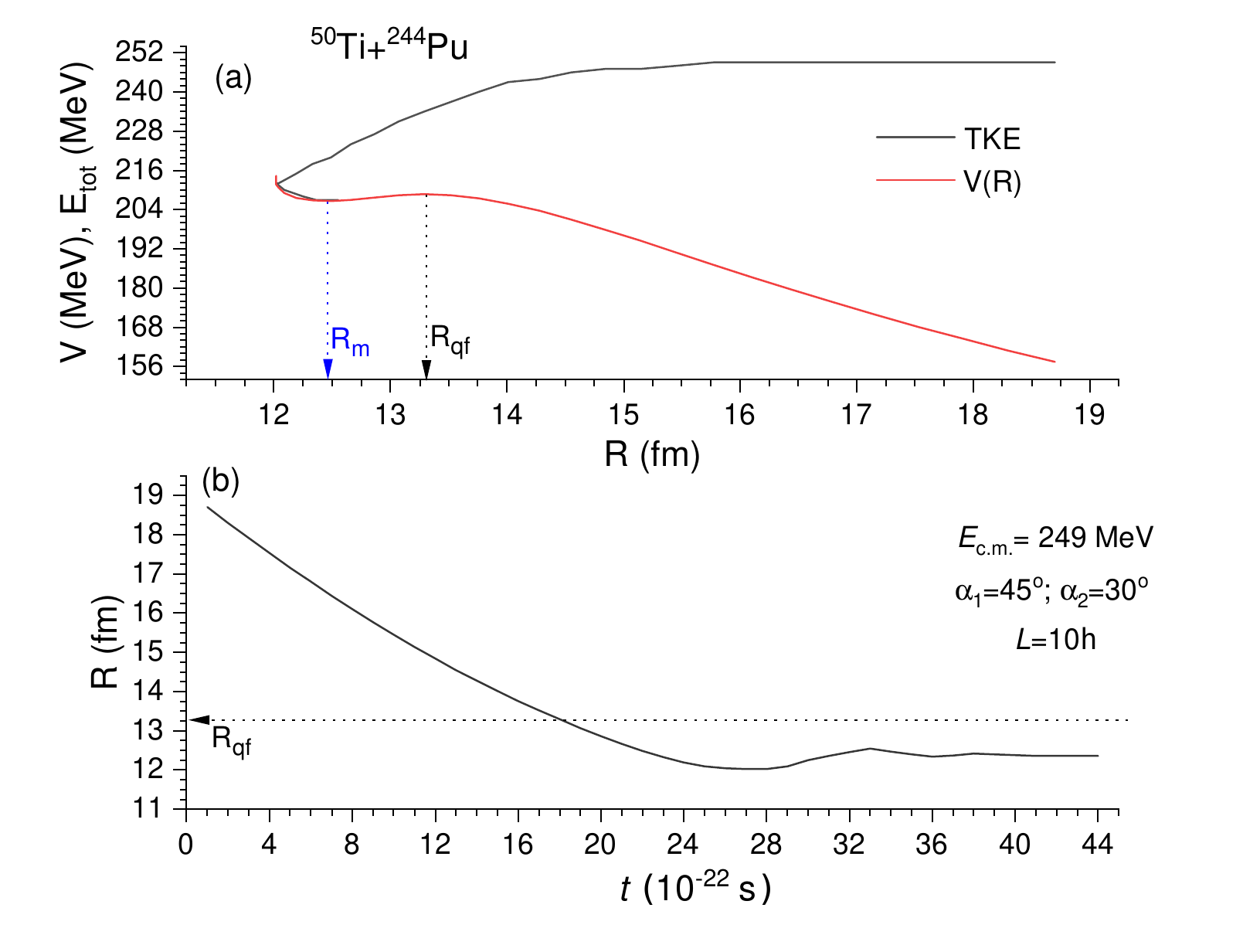}
\caption{(Color online) Trajectory of the capture for $^{50}$Ti$+^{244}$Pu reaction at energy $E_{c.m.}=249$MeV, angular momentum $L=10\hbar$,  orientation angles of the axial symmetry axis have $45^\circ$ and $30^\circ$.}
\label{cap}
\end{figure}

In deep-inelastic collisions, full momentum transfer of the relative motion does not occur, and the interaction time between the colliding nuclei is relatively short compared to capture reactions, where full momentum transfer is required. In capture reactions, the total kinetic energy of the products is fully dissipated, resulting in significantly lower values than the initial collision energy. 
Capture occurs when two conditions are satisfied: 
\begin{itemize}
\item The initial projectile energy in the center-of-mass frame must be sufficient to reach the potential $V(R,\ell,\alpha_1,\alpha_2)$ well of the nucleus-nucleus interaction, by overcoming or tunneling through the entrance-channel barrier.
\item  At the same time the value of the dissipated kinetic energy  should correspond with the size of the potential well.
\end{itemize} 
Conversely, in deep-inelastic collisions, the total kinetic energy is not entirely damped, and its value remains closer to the initial values. In the DNS approach the difference between capture and deep-inelastic collisions depends on whether the path of relative motion has been trapped into the potential well or not. 

\begin{figure}[ ]
  \centering
  \includegraphics[width=0.5\textwidth]{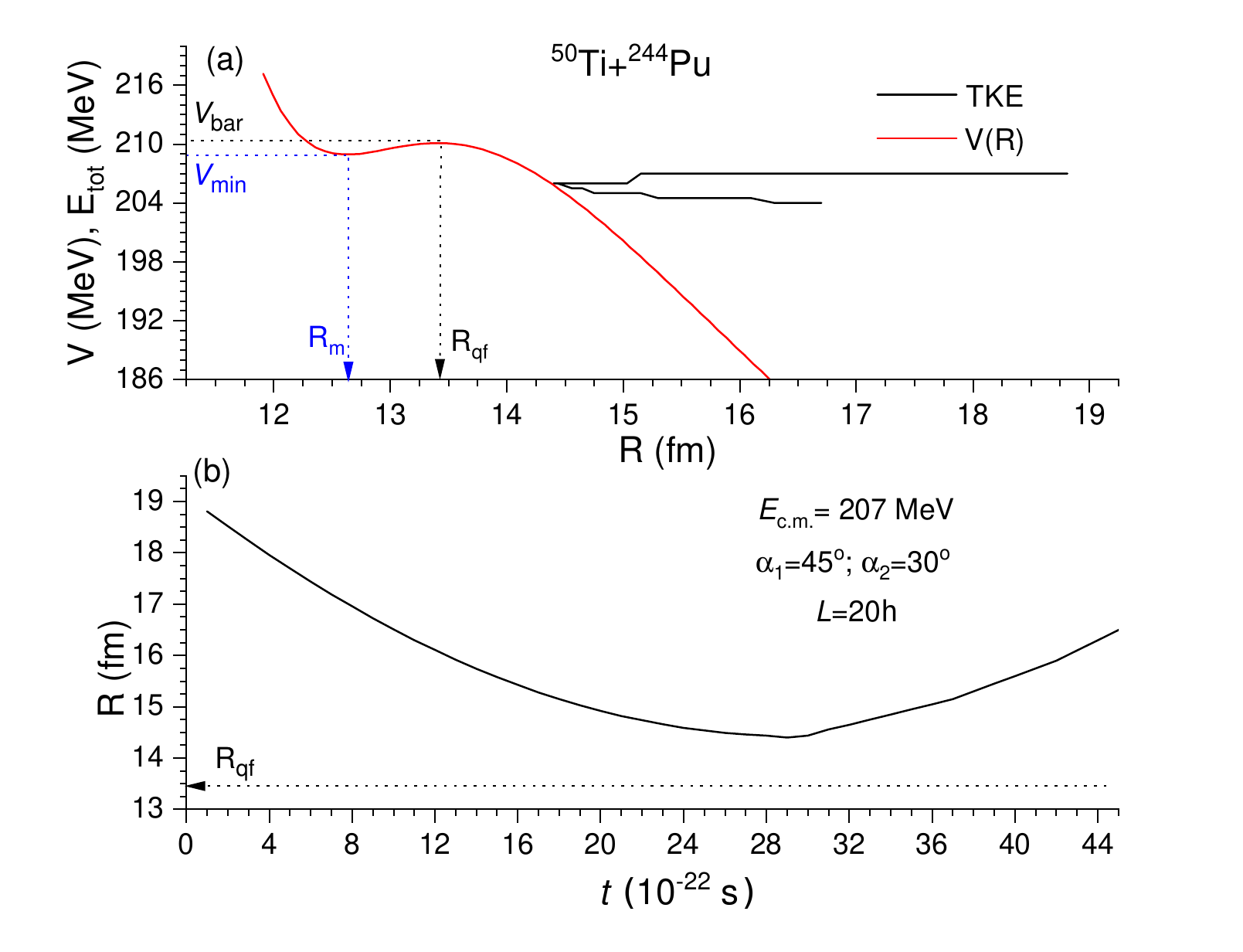}
\caption{(Color online) Trajectory of the motion when there is no capture for $^{50}$Ti$+^{244}$Pu reaction at energy $E_{c.m.}=207$MeV, angular momentum $L=20\hbar$, orientation angles of the axial symmetry axis have $45^\circ$ and $30^\circ$.}
\label{nocapture}
\end{figure}

Figure \ref{cap}(a) illustrates the dissipation of kinetic energy during the evolution of relative motion. As the system overcomes the interaction barrier, the kinetic energy is fully dissipated, ultimately settling at the bottom of the potential well. In Figure \ref{cap}(b), the solution of equation (\ref{maineq}) is presented, showing the variation of the relative distance between the nuclei over time at an energy of 249 MeV for the $^{50}$Ti$+^{244}$Pu reaction. It can be seen that the relative distance between the nuclei decreases until it reaches a value approximately equal to the sum of their radii, after which it remains constant over time. This indicates that the projectile nucleus has been successfully captured by the target nucleus. 

Figure \ref{nocapture}(a) corresponds to the trajectory of deep-inelastic collisions at center-of-mass energy $E_{\rm c.m.}$=207 MeV. In this case, full momentum transfer of the relative motion does not occur, preventing the formation of a compound system. It is evident from Figure \ref{nocapture}(a) that at this energy, the projectile is unable to overcome the barrier. As a result, the relative distance between the nuclei begins to increase before reaching the barrier position (Fig. \ref{nocapture}(b)), preventing the capture process.

By utilizing key reaction parameters such as the center-of-mass energy  $E_{\rm c.m.}$ and angular momentum $\ell$, which determine the conditions for capture, the theoretical capture cross sections can be estimated using the following formula \cite{Kayumov2022}:
\begin{eqnarray}\label{capture}
\sigma_{\rm cap}(E_{\rm c.m.},\ell,\alpha_i)&=&\frac{\lambda^2}{4\pi}(2\ell+1)\nonumber\\
&\times&\mathcal{P}^{\ell}_{\rm cap}(E_{\rm c.m.},\alpha_i),
\end{eqnarray}
where $\mathcal{P}^{\ell}_{cap}(E_{\rm c.m.},\ell,\alpha_i)$ is the capture probability, $\lambda$ is the de Broglie wavelength of the entrance channel, $\alpha_i$ are their orientation angles of the axial symmetry axis for projectile and target nuclei, $i=1$ and 2 corresponding. The calculation of the capture probability is based on the following condition: 
\begin{widetext}
\begin{eqnarray}
\mathcal{P}^{(\ell)}_{cap}(E_{\rm c.m.},\alpha_i) = \left\{
\begin{array}{lll}
1,\,\, \mbox{if}\,\, \ell_m < \ell < \ell_d \,\, \mbox{and} \,\, E_{\rm c.m.} > V_{B}, \\
0,\,\, \mbox{if} \,\,\ell < \ell_m \,\, \mbox{or}\,\,\ell>\ell_d \,\, \mbox{and} \,\, E_{\rm c.m.}>V_{B}\, \\
\mathcal{P}^{(\ell)}_{tun}(E_{\rm c.m.},\alpha_i),\,\, \mbox{for all}\,\, \ell \mbox{ if } V_{\rm min} < E_{\rm c.m.}\leq V_{B},\, \\
\end{array} \right.
\label{CapClass}
\end{eqnarray}
\end{widetext}
here, $V_B$ represents the barrier height, while $V_{\rm min}$ denotes the minimum of the potential well in the nucleus-nucleus interaction (see Fig.\ref{nocapture}). The boundary values  $\ell_m$  and $\ell_d$ are determined by solving the equation for the relative motion (\ref{maineq}), defining the range of angular momentum values ($\ell$-window) that contribute to the capture process. The probability of barrier penetration, $\mathcal{P}^{(\ell)}_{\rm tun}$, is evaluated using the improved WKB approximation, as represented by Kemble (1935) \cite{Kemble1935}:

\begin{eqnarray}\label{penetration}
\mathcal{P}^{(\ell)}_{tun}(E_{\rm c.m.},\alpha_i)=\frac{1}
{1+\exp\left[2K(E_{\rm c.m.},\ell,\alpha_i)\right]},\nonumber\\
\end{eqnarray}
where
\begin{eqnarray}
&&K(E_{\rm c.m.},\ell,\alpha_i)\nonumber\\
&&=\int\limits_{R_{in}}^{R_{out}} dR\times\sqrt{\frac{2\mu}{\hbar^2}(V(R,\ell,\alpha_i)-E_{\rm c.m.})}.
\end{eqnarray}

The inner and outer turning points, denoted as $R_{in}$ and $R_{out}$, correspond to the radial distances at which the nucleus-nucleus interaction potential equals the collision energy in the center-of-mass frame $V(R,\ell,\alpha_i)=E_{\rm c.m.}$.

\section{Compound nucleus formation probability}
The formation of a compound nucleus (CN) is a crucial stage in heavy-ion collisions, determining the success of fusion reactions, particularly in the synthesis of superheavy elements (SHEs). In the DNS model, the probability of CN formation is governed by the competition between fusion, quasifission, and fast-fission, where the interacting nuclei may either evolve into a fully equilibrated compound system or re-separate before fusion occurs. Within the DNS approach, the CN probability, is determined by the competition between these three processes and is typically obtained by solving a master equation that describes the evolution of the mass and charge distributions of the DNS \cite{Giardina2000, Nasirov2005, Kayumov2022}:
\begin{eqnarray}
P_{\rm CN}(E_{\rm c.m.},\ell,\alpha_i)
&=&\sum\limits^{Z_{max}}_{Z_{sym}}D_Z(E^*_Z,\ell,\alpha_i)\nonumber\\
&\times& 
P^{(Z)}_{\rm CN}(E^*_Z,\ell,\alpha_i),
\label{PcnDZ}
\end{eqnarray}
where 
\begin{eqnarray}
E^*_Z(\ell,\alpha_i)&=&E_{\rm c.m.}-V_{\rm min}(Z,A,R_m,\alpha_i)\nonumber\\
&-&\Delta Q_{gg}(Z,A),
\label{ExiZ}
\end{eqnarray}

and change in binding energies is $\Delta Q_{gg}(Z,A)=B_P(Z_P,A_P)+B_T(Z_T,A_T)-(B_1(Z,A)+B_2(Z_c,A_c))$ which is calculated by using the table values from Ref. \cite{AUDI2003, MolNix1995}, $V_{\rm min}(Z,A,R_m,\alpha_i)$ is potential energy at $R=R_m$ for the corresponding $Z$ and $A$. 
The values of $D_Z(E^*_Z,\ell,\alpha_i)$ are obtained by solving the transport master equation, where the nucleon transition coefficients depend on the occupation numbers and the single-particle energies of the nucleons in the DNS nuclei (detailed calculation was provided in \cite{Kayumov2022}). The fusion probability, $P^{(Z)}_{CN}(E^*_Z,\ell,\alpha_i)$, for DNS fragments with charge configuration $Z$  and orbital angular momentum $\ell$ is calculated as the branching ratio of the level density functions of the Fermi system associated with the quasifission barrier, $B^{(Z)}_{qf}(\ell,\alpha_i)$, at a given mass asymmetry to those corresponding to the intrinsic fusion barrier, $B^{*(Z)}_{fus}(\ell,\alpha_i)$, and the symmetry barrier, $B^{*(Z)}_{\rm sym}(\ell,\alpha_i)$, along the mass asymmetry axis \cite{Nasirov2019}:

\begin{widetext}
\begin{equation}
\label{Pcn} 
P^{(Z)}_{CN}((E^*_Z,\ell,\alpha_i))=\frac{e^{-B_{\rm fus}^{*(Z)}(\ell,\alpha_i)/T_Z}}{e^{-B_{\rm fus}^{*(Z)}(\ell,\alpha_i)/T_Z} + e^{-B_{\rm qf}^{(Z)}(\ell,\alpha_i)/T_Z}+e^{-B_{\rm sym}^{*(Z)}(\ell,\alpha_i)/T_Z}}.
\end{equation}
\end{widetext}

The effective temperature of the DNS $T_Z=T_Z(\ell,\alpha_i)$, corresponding to the charge number of the light fragment, can be expressed as follows:
\begin{equation}
   T_Z(\ell,\alpha_i)= \sqrt{\frac{12E^*_Z(\ell,\alpha_i)}{A_{\rm CN}}}.
\end{equation}

The values for the barriers $B_{\rm fus}^{*(Z)}(\ell,\alpha_i)$, $B_{\rm sym}^{*(Z)}(\ell,\alpha_i)$ can be calculated by using driving potential, and value for the quasifission barrier estimated by $B_{\rm qf}^{(Z)}(\ell,\alpha_i)=V_{\rm bar}(Z,A,R_{\rm qf},\alpha_i)-V_{\rm min}(Z,A,R_{\rm m},\alpha_i)$ (see Fig.\ref{nocapture} (a)). 
On the other hand, the calculation of the driving potential involves analyzing the mass asymmetry and potential energy surface (PES) which is calculated by using fragments' mass and charge numbers, orbital angular momentum of the system.   
\begin{eqnarray}\label{pes}
U(Z,A,\ell,R,\alpha_i)&=&V(Z,A,\ell,R,\alpha_i)\nonumber\\
&+&Q_{gg}-V_{rot}^{CN}(\ell),
\end{eqnarray}
the way how the potential $V(Z_i,A_i,\ell,R,\alpha_i)$ was calculated is presented in Appendix A of Ref. \cite{Nasirov2005}, $Q_{gg}=B_P(Z_P,A_P)+B_T(Z_T,A_T)-B(Z_{\rm CN},A_{\rm CN})$ is the reaction balance energy, $V_{rot}^{CN}(\ell)$ is the rotational energy of CN. The curve connecting the minima along the potential energy surface (PES) is used as the driving potential, for the DNS formed in the given reaction it follows:
\begin{eqnarray}
U_{\rm dr}(Z,A,\ell,\alpha_i) &=& V(Z,A,\ell,R_{\rm m}, 
\alpha_i)+Q_{gg} \cr
&-&V_{rot}^{CN}.
\label{driving}
\end{eqnarray}

\begin{figure}[ ]
\includegraphics[width=0.50\textwidth]{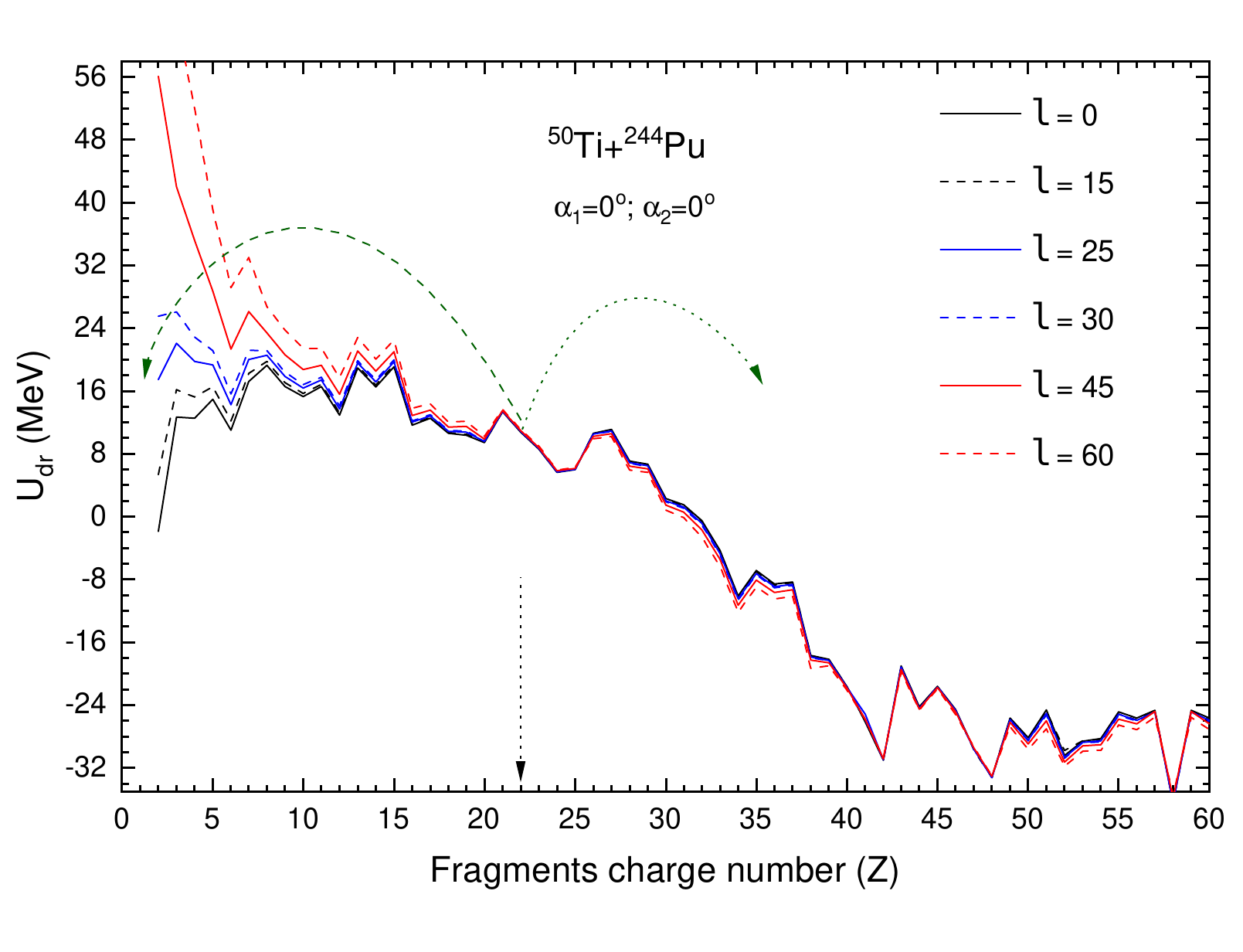}
\caption{(Color online) The driving potential for the DNS formed in the $^{50}$Ti+$^{244}$Pu reaction calculated for different angular orbital momentum values. The fusion direction as a next step for the evolution of DNS is shown with green solid line, for the charge asymmetry of the entrance channel $Z=22$.}
\label{driving116}
\end{figure}

In figure \ref{driving116} results for the driving potential presented for the reaction $^{50}$Ti+$^{244}$Pu at different values of orbital angular momentum. In the direction of the evolution of DNS to compound nucleus, which is presented with green solid line in the figure, there is a barrier which corresponds the value where $Z_{bar}=15$ for $L=0\hbar, 15\hbar, 25\hbar$. This value can be calculated easily by using $B_{\rm fus}^{*(Z)}(\ell,\alpha_i)=U_{\rm dr}(Z_{bar},A_{bar},\ell,\alpha_i)-U_{\rm dr}(Z_P,A_P,\ell,\alpha_i)$ expression. It is clear from the figure that the fusion barrier value $B_{\rm fus}^{*(Z)}(\ell,\alpha_i)$ rises with increasing angular orbital momentum, which indicates that a DNS with higher rotational energy becomes less favorable to forming the compound nucleus. 

\begin{figure}[ ]
\includegraphics[width=0.50\textwidth]{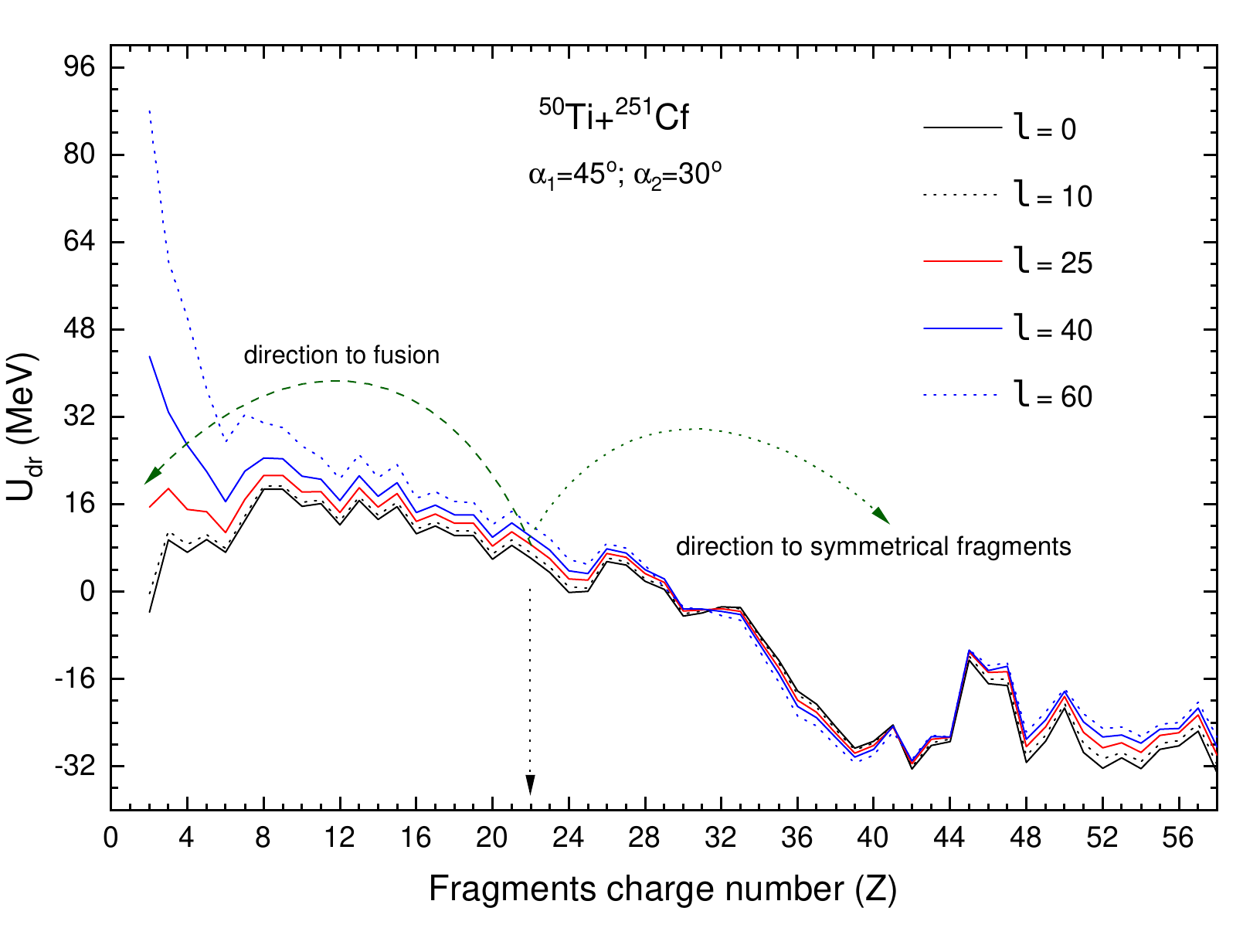}
\caption{(Color online) Same as figure \ref{driving116}, but for  $^{50}$Ti+$^{251}$Cf reaction at $\alpha_1=45^\circ; \alpha_2=30^\circ$ orientation angles of the axial symmetry axis.} 
\label{driving120}
\end{figure}

Figure \ref{driving120} represents result of calculation of the driving potential for $^{50}$Ti+$^{251}$Cf reaction for the different values of angular orbital momentum. The fusion barrier is significantly lower when angular momentum 0$\hbar$-25$\hbar$ and its position located at $Z=9$. For the values 40$\hbar$ and above, the fusion barrier increases dramatically, which reduces the fusion probability of the DNS and by increasing probability the breakup of the system into two fragments.  

\begin{figure}[ ]
\includegraphics[width=0.50\textwidth]{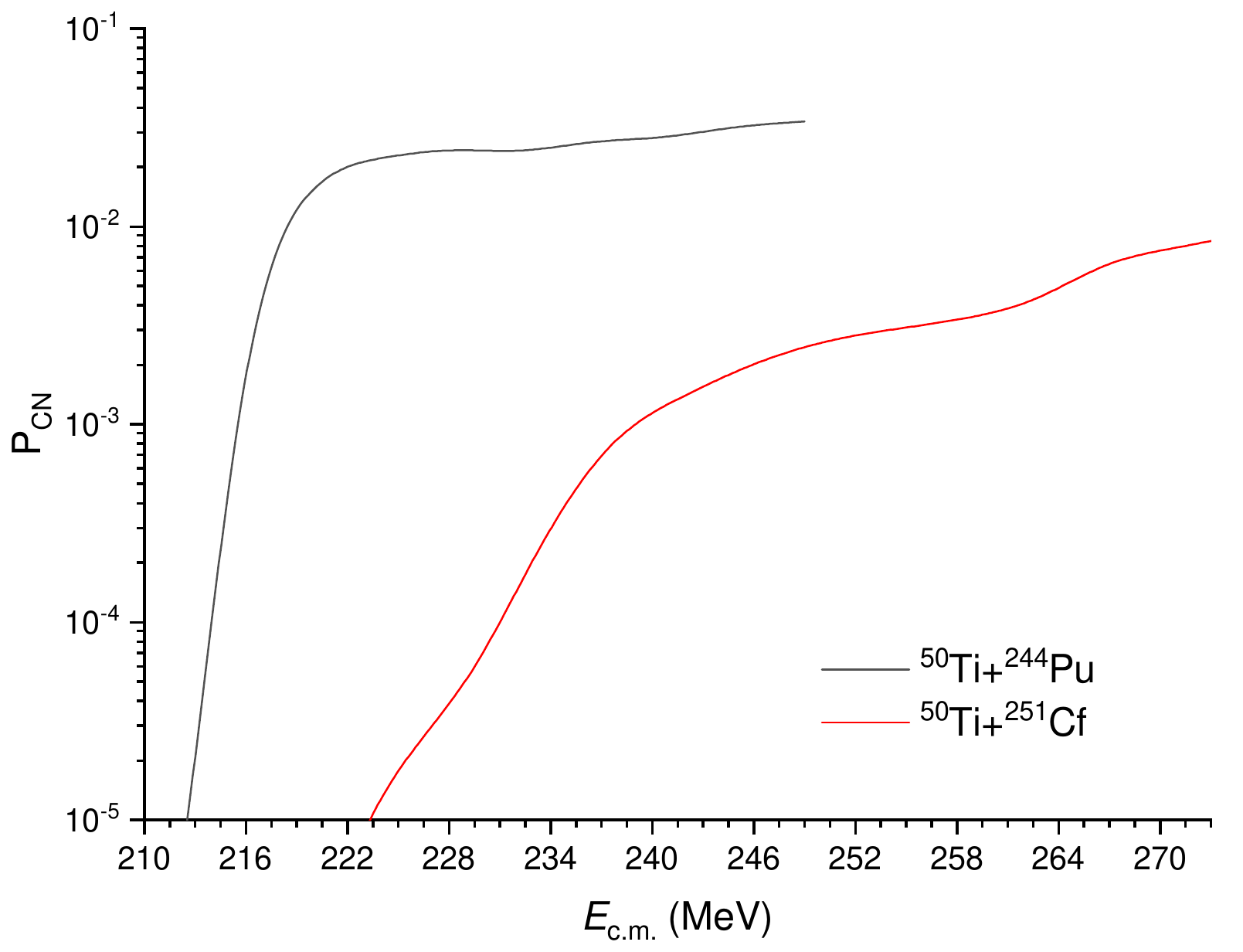}
\caption{(Color online) Dependence of the probability of formation of a compound nucleus ($P_{\rm CN}$) on the collision energy in center of mass system for the reactions $^{50}$Ti+$^{244}$Pu and $^{50}$Ti+$^{251}$Cf.} 
\label{CompPcn}
\end{figure}

Estimated values of the $B_{\rm fus}^{*(Z)}(\ell,\alpha_i)$ and $B_{\rm sym}^{*(Z)}(\ell,\alpha_i)$ for these reactions, can be used to promote calculation of the fusion probability $P^{(Z)}_{CN}(E^*_Z,\ell,\alpha_i)$ for DNS fragments with charge configuration $Z$  and orbital angular momentum $\ell$ by using (\ref{Pcn}). Overall, by employing equation (\ref{PcnDZ}), the fusion probability of the colliding nuclei is taken as a sum over different charge asymmetry from the configuration starting from $Z_{sym}=\frac{Z_P+Z_T}{2}$ up to $Z_{max}=Z_P+Z_T$ and the results is presented in Fig.\ref{CompPcn} for the reactions $^{50}$Ti+$^{244}$Pu and $^{50}$Ti+$^{251}$Cf. 

\section{Fusion and Evaporation residue cross sections}
The partial fusion cross section is determined by the product of capture cross section which was calculated in equation (\ref{capture}) and CN probability from equation (\ref{PcnDZ}) for different values of the orientation angles of the axial symmetry axis and for the various excitation energies:   
\begin{eqnarray}
    \sigma_{\rm fus}(E_{\rm c.m.}, \ell, \alpha_i)&=&
    \sigma_{\rm cap}(E_{\rm c.m.}, \ell, \alpha_i)\cr
    &\times& P_{CN}(E_{\rm c.m.}, \ell, \alpha_i).
    \label{fusl}
\end{eqnarray}

\begin{figure}[b]
\includegraphics[width=0.50\textwidth]{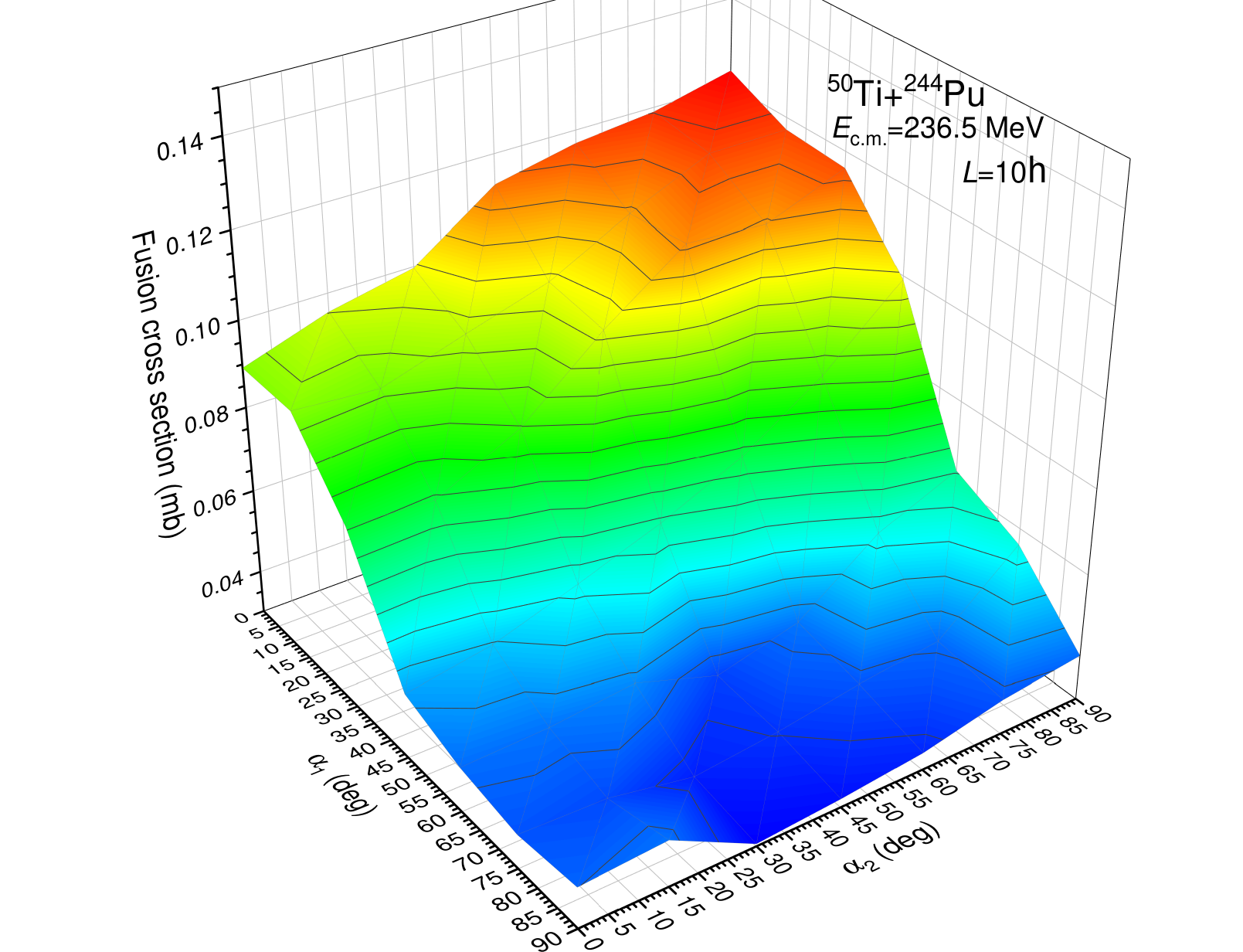}
\caption{(Color online) Fusion cross section as orientation angles function for the $^{50}$Ti+$^{244}$Pu reaction, at center-of-mass energy $E_{\rm c.m.}$=236.5 MeV and at the value of angular momentum 10$\hbar$.}
\label{angfusion116}
\end{figure}

\begin{figure}[ ]
\includegraphics[width=0.50\textwidth]{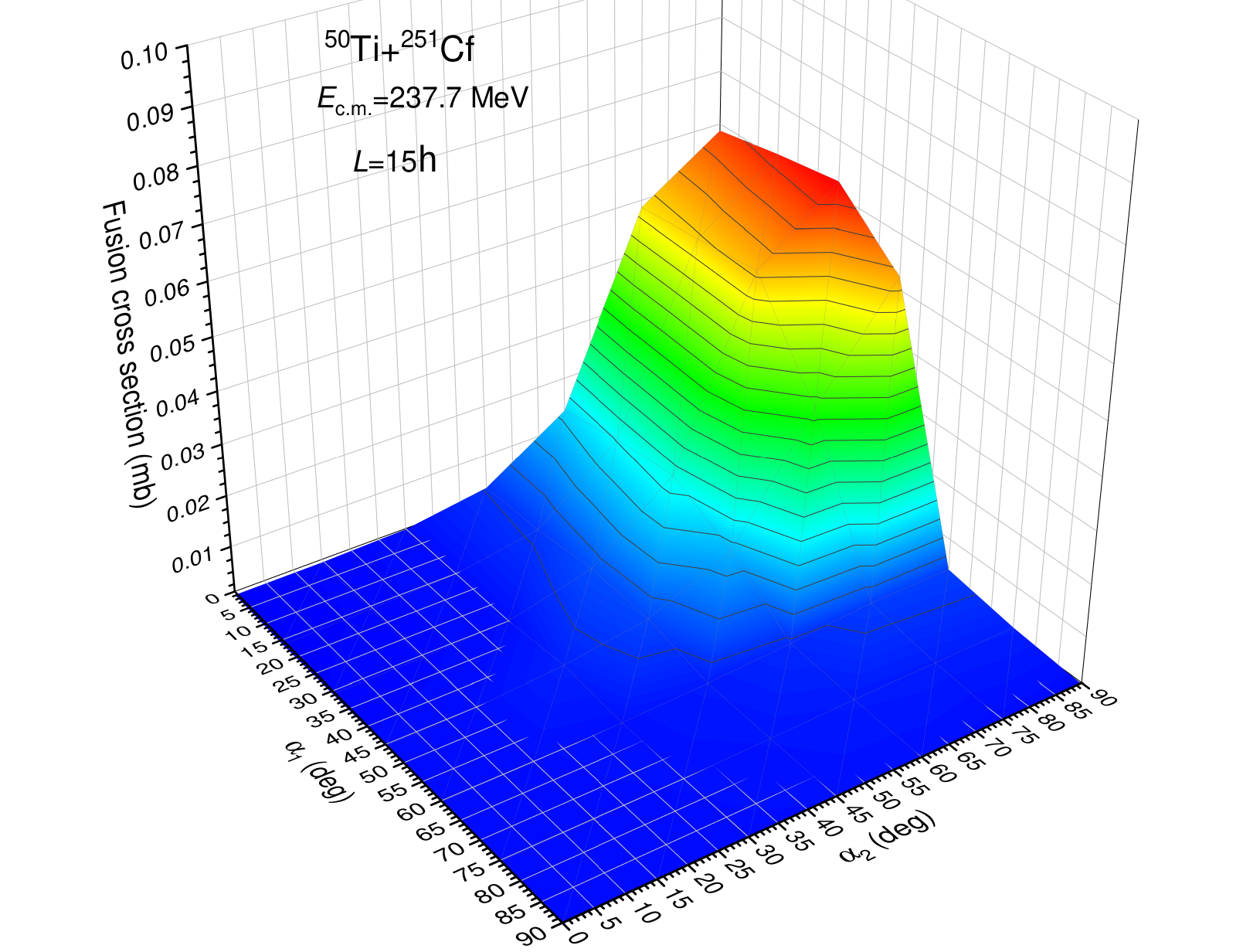}
\caption{(Color online) Same as Fig.\ref{angfusion116} for the $^{50}$Ti+$^{251}$Cf reaction at $E_{\rm c.m.}$=237.7 MeV and 15$\hbar$.} 
\label{angfusion120}
\end{figure}

Figures \ref{angfusion116} and \ref{angfusion120} represents how the fusion cross section differs for the various values of the orientation angles for projectile $\alpha_1$ and target $\alpha_2$ nuclei. It can be seen from figure \ref{angfusion116} that the maximum value of fusion cross section correspond when $\alpha_1=0^\circ$ and  $\alpha_2=90^\circ$ (equatorial collisions), where other values is distributed by all other possible orientational angles for $^{50}$Ti+$^{244}$Pu reaction at $E_{\rm c.m.}$=236.5 MeV. This specific configuration likely corresponds to an optimal overlap between the interacting nuclear surfaces, thereby lowering the effective fusion barrier and enhancing the probability for compound nucleus formation. 
However, fusion cross section for the $^{50}$Ti+$^{251}$Cf reaction reaches the maximum around $\alpha_1=30^\circ-55^\circ$ and $\alpha_2=90^\circ$ (Fig. \ref{angfusion120}), for all values $\alpha_2<35^\circ$ there is no fusion. For the Cf target, the tip-on (i.e. smaller $\alpha_2$) configuration does not provide sufficient overlap or may result in an increase in the fusion barrier due to dependence of the nucleus-nucleus potential $V(R,\ell,\alpha_1,\alpha_2)$  on the geometric configuration of colliding nuclei. Overall, these results highlight the role of the  orientation angles of nuclei in calculation of the fusion process and have well agreement with the experimental observations and conclusions \cite{Nishio2000, Oganessian2004PRC, Nishio2012}.

\begin{table}[ht]
\caption{Deformation parameters which are used for the ground states $\beta_2$ and $\beta_3$ and for the first excited  2$^+$ and $3^-$ states.}
\begin{ruledtabular}
\begin{tabular}{ccccc}
Nucleus & $\beta_2$ \cite{MolNix1995}  & $\beta_3$ \cite{MolNix1995} &
$\beta_{2+}$ \cite{Raman2001} & $\beta_{3-}$ \cite{Spear1989} \\
\hline
$^{50}$Ti & 0.0 & 0.0 & 0.230 &  0.16\\
$^{244}$Pu & 0.237 & 0.061  & 0.289 & 0.090\\
$^{251}$Cf & 0.250 & 0.027  & 0.299 & 0.050\\
\end{tabular}
\end{ruledtabular}
\label{defpara}
\end{table}

Parameters for the quadrupole and octupole deformations of the ground states of the reacting nuclei are taken from Ref. \cite{MolNix1995}. The deformation parameters for the first excited $2^+$ and $3^-$ states are taken from Refs. \cite{Raman2001} and \cite{Spear1989}, respectively. The target nuclei in the both reactions, $^{244}$Pu and $^{244}$Pu, exhibits a deformed ground state (see Table \ref{defpara}). In the case of the projectile nucleus, $^{50}$Ti, is spherical in its ground state (see Table \ref{defpara}); however, its first excited quadrupole state, with \(\beta_{2+} = 0.23\), is treated as a zero-point vibrational state. The orientation angle of the spherical projectile, $\alpha_1$, is set to $0^\circ$ for simplicity. Consequently from these, the partial fusion cross section is determined by averaging over the vibrational state parameters $\beta_2$ and $\beta_3$ of the projectile and over different orientation angles $\alpha_2$ of the target nucleus relative to the beam direction.

\begin{eqnarray}
\langle \sigma_{\rm fus} (E_{\rm c.m},\ell) \rangle &=& 
\int^{\beta_{2+}}_{-\beta_{2+}}d\beta^{(1)}_2\int^{\beta_{3-}}_{-\beta_{3-}}d\beta^{(1)}_3 g(\beta^{(1)}_{2},\beta^{(1)}_{3})
\nonumber\\
&\times& \int_0^{\pi /2} \sigma_{fus}(E_{\rm c.m},\ell,\alpha_2)\sin\alpha_2 d\alpha_2 
\label{vibr}
\end{eqnarray}
where surface vibrations of the projectile nucleus are used as independent harmonic oscillators, and the nuclear radius is taken into account as a Gaussian distribution \cite{Esbensen1981}:

\begin{eqnarray}
g(\beta_2,\beta_3) = \exp
\left[ -\frac{R_0^2\left[\sum_{\lambda}\beta_{\lambda} Y_{\lambda0}^* (\alpha_1)\right]^2}{2 \sigma_{\beta}^2} \right] (2\pi \sigma_{\beta}^2)^{-1/2}.
\end{eqnarray}
\begin{eqnarray}
\sigma^2_{\beta_2} = \frac{R_0^2}{4\pi}\sum_{\lambda} \beta_{\lambda}^2.
\end{eqnarray}

\begin{figure}[t]
\includegraphics[width=0.50\textwidth]{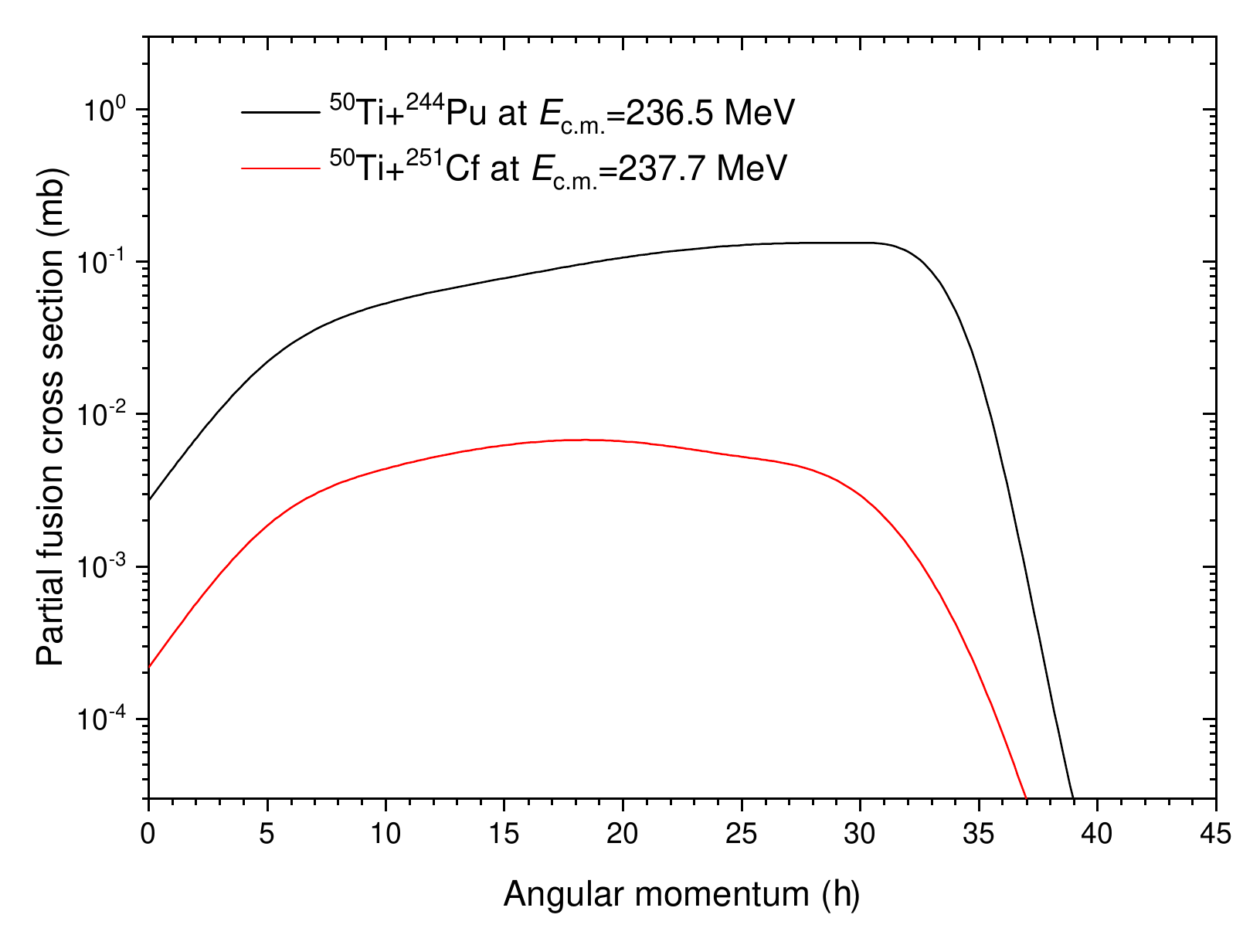}
\caption{(Color online) Fusion cross section as a function of angular momentum for the $^{50}$Ti+$^{244}$Pu and $^{50}$Ti+$^{251}$Cf reactions, at center-of-mass energy $E_{\rm c.m.}$=236.5 MeV and $E_{\rm c.m.}$=237.7 MeV, respectively.}
\label{parfus}
\end{figure}

\begin{figure}
\includegraphics[width=0.50\textwidth]{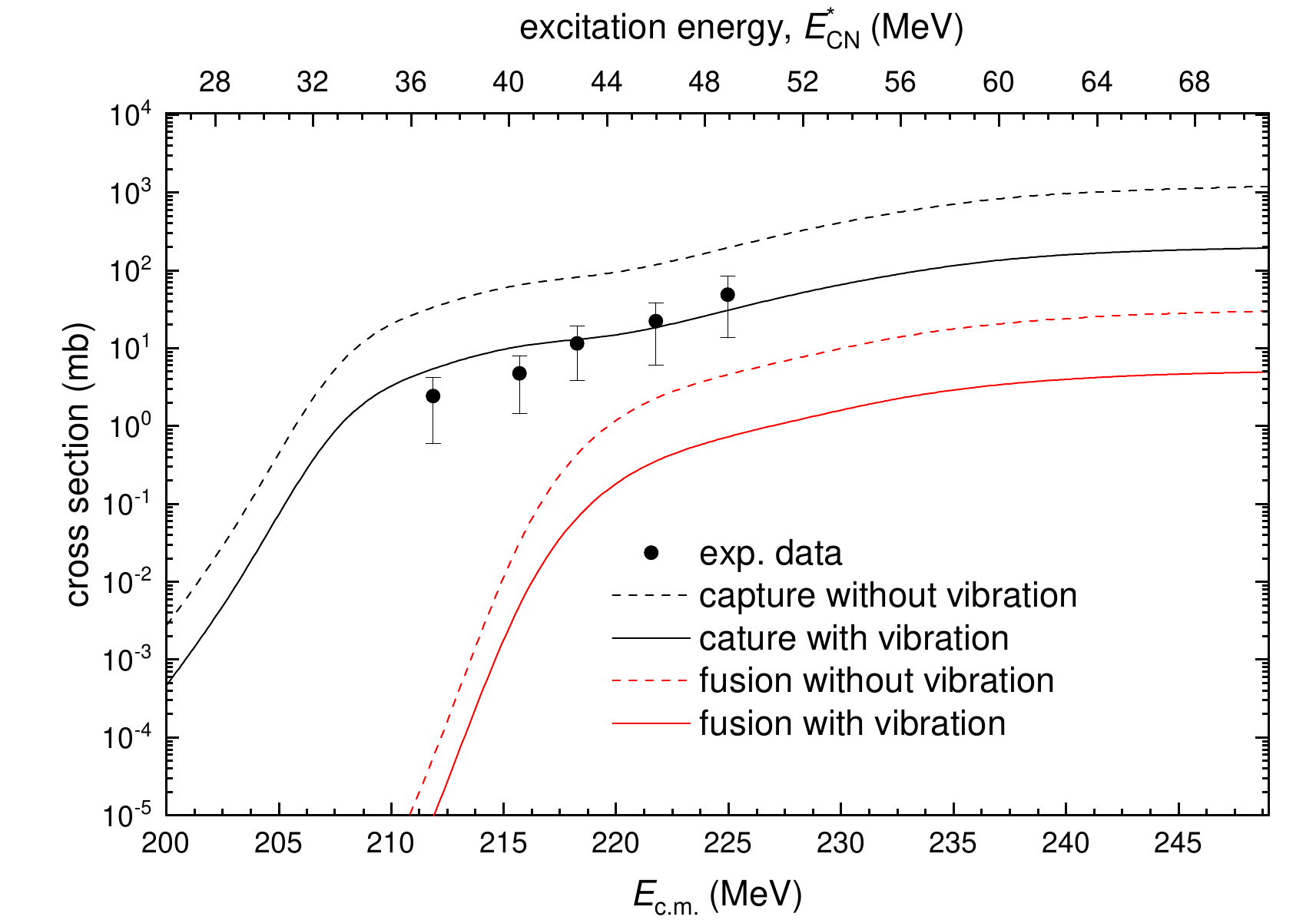}
\caption{(Color online) Comparison of calculated and experimental data of cross sections for the capture (black curves) and the complete fusion (red curves) for the $^{50}$Ti+$^{244}$Pu reaction. Experimental data for the capture cross section are taken from \cite{ITKIS2007150}.}
\label{capture116}
\end{figure}

\begin{figure}[ ]
\includegraphics[width=0.50\textwidth]{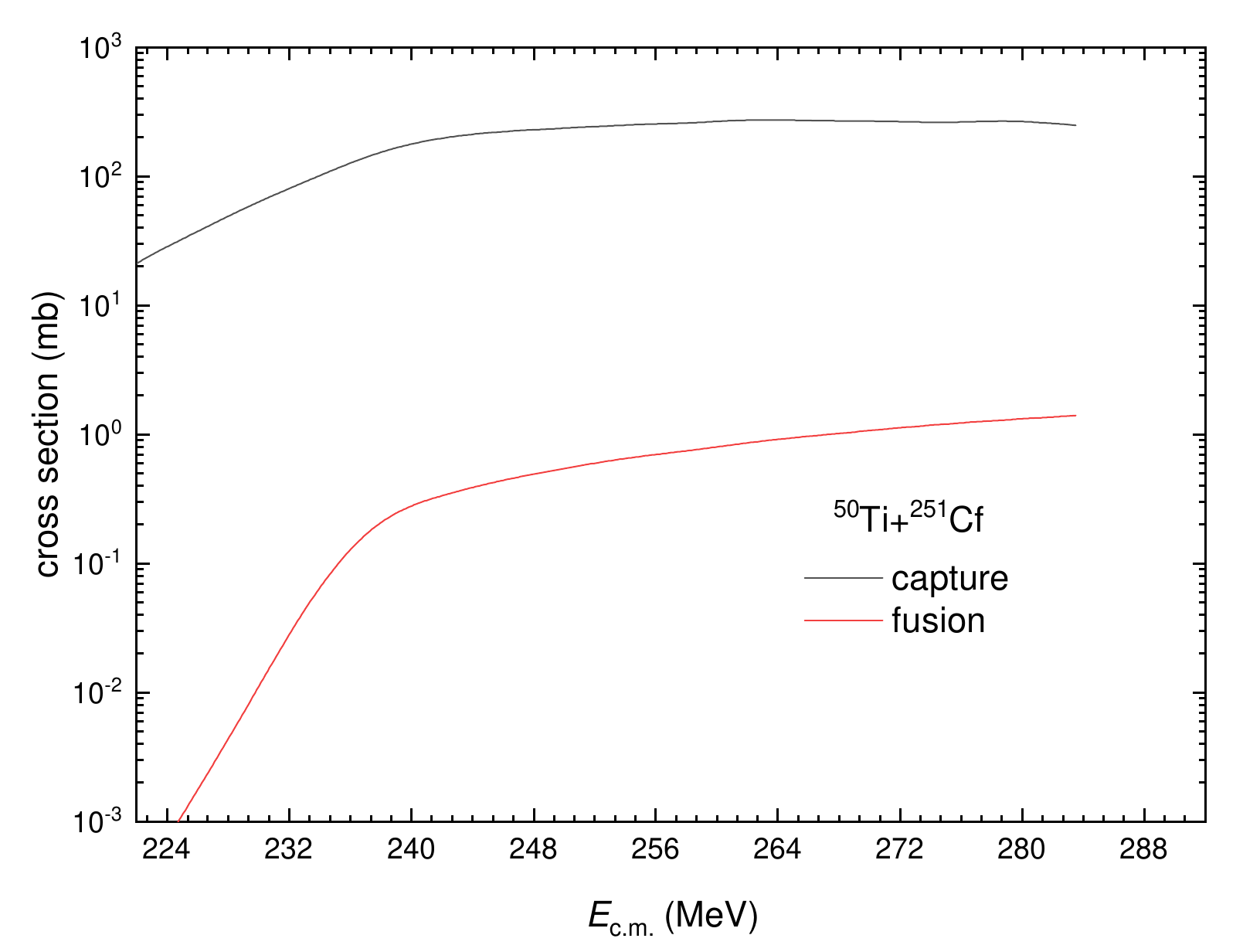}
\caption{(Color online) Same as Fig. \ref{capture116}, but for the $^{50}$Ti+$^{251}$Cf reaction by taking into account averaging over the vibrational state parameters.}
\label{capture120}
\end{figure}

As a result of these calculation partial fusion cross section can be written as $\sigma_{\rm fus} (E_{\rm c.m},\ell)=\langle \sigma_{\rm fus} (E_{\rm c.m},\ell) \rangle$ and it is represented in figure \ref{parfus}. It should be noted that the averaging procedures over orientation angles ($0^\circ, 15^\circ, 30^\circ, 45^\circ, 60^\circ, 75^\circ$, and $90^\circ$) of the axial symmetry axis of the deformed  nuclei and over 7 vibrational states ($-\beta^+_2, -2\beta^+_2/3, -\beta^+_2/3, 0, \beta^+_2/3, 2\beta^+_2/3, \beta^+_2$) of the spherical nucleus $^{50}$Ti have been performed to get the partial cross sections in Fig. \ref{parfus}. 
Overall, the total fusion cross section can be obtained by summing the contributions from all partial waves:
\begin{eqnarray}  \label{compfus}
\sigma_{\rm fus}(E_{\rm c.m.})=\sum_{\ell}\sigma_{\rm fus} (E_{\rm c.m},\ell)&=&\sum_{\ell}\sigma_{\rm cap}(E_{\rm c.m.},\ell)\nonumber\\ &\times& P_{\rm CN}(E_{\rm c.m.},\ell).
\end{eqnarray}
Results of this calculation are presented in Figs. \ref{capture116} and \ref{capture120}, for the $^{50}$Ti+$^{244}$Pu and $^{50}$Ti+$^{251}$Cf reactions, correspondingly. 

Figure \ref{capture116} shows the calculated capture and fusion cross sections for the $^{50}$Ti+$^{244}$Pu reaction as a function of the center of mass energy $E_{\rm c.m.}$. In figure dotted lines correspond to the averaging only over the different orientation angles $\alpha_2$ of the target, whereas the solid curves presents results of calculation by additionally averaging over the vibrational state parameters $\beta_2$ and $\beta_3$ of the projectile according to Eq.(\ref{vibr}). Calculated capture cross section with averaging over the vibrational state are in good agreement with the experimental data \cite{ITKIS2007150}. While the estimates that exclude vibrational state averaging provided higher values for cross sections.
Figure \ref{capture120} presents the capture and the fusion cross sections for the $^{50}$Ti+$^{251}$Cf reaction which are calculated by averaging over vibrational states for projectile nucleus and orientation angles for target nucleus by (\ref{vibr}). The maximum fusion probability for the $^{50}$Ti+$^{251}$Cf system is achieved at a different energy compared to the $^{50}$Ti+$^{244}$Pu reaction. This difference is primarily due to the higher charge and deformation effects of the Cf target, which changes the nucleus–nucleus interaction potential and shift the effective fusion barrier. Together, these figures underscore the critical influence of the colliding nuclei's structural properties on the fusion process and highlight the importance of selecting appropriate reaction conditions for synthesizing superheavy elements.

Evaporation residue cross sections for the emission of $x$ neutrons at a given energy and orbital angular momentum were calculated using \cite{Nasirov2005, Mandaglio2012}:
\begin{eqnarray}\label{ercs}
\sigma^x_{ER}(E^*_{x},\ell)=\sigma^{x-1}_{ER}(E^*_{x-1},\ell)W^x_{\rm sur}(E^*_{x-1},\ell).
\end{eqnarray}
where, as the energy excitation energy of the system $E^*_{x}$ is used; $W^x_{\rm sur}(E^*_{x-1},\ell)$ represents the survival probability of the $x$th intermediate nucleus against fission during the de-excitation cascade of the compound nucleus. This survival probability is calculated using the statistical model implemented in KEWPIE2 \cite{Kewpie2}. $\sigma^{x-1}_{ER}(E^*_{x-1},\ell)$ denote the partial cross section for forming the intermediate excited nucleus at the $(x-1)$th step and obviously, when $x=1$: 
\begin{eqnarray}\label{ercs}
\sigma^{(x-1)}_{ER}(E^*_{x-1},\ell)=\sigma^{(0)}_{ER}(E^*_{0},\ell)=\sigma_{\rm fus}(E^*_{CN},\ell),
\end{eqnarray}
and results of the calculated partial cross section (Eq.(\ref{vibr})) can be used for $x=1$.  
In the calculation of the survival probability $W^x_{\rm sur}(E^*_{x-1},\ell)$, the Weisskopf-Ewing model \cite{Weisskopf} and the standard Bohr–Wheeler transition-state model \cite{Wheeler} are used to calculate neutron emission width and the fission-decay width, correspondingly (details of the calculation provided here \cite{Nasirov2024,Nasirov2024232TH}). Additionally, when estimating the fission-decay width, the effect of viscosity in the fission process accounted for by the Kramers correction factor \cite{Kramers1940}, and the difference in the number of stationary collective states between the ground state and the saddle point incorporated via the Strutinsky factor \cite{STRUTINSKY1973} were taken into account. 
In calculating the survival probability $W^x_{\rm sur}(E^*_{x-1},\ell)$, the dependence of the fission barrier $B_f$ on the compound nucleus excitation energy $E^*_{\rm CN}$ is incorporated using the relation
\begin{eqnarray}
B_f = B_{\rm LSD} - f\,\delta W,
\end{eqnarray}
where $B_{\rm LSD}$ is the empirical fission barrier determined by the Lublin-Strasbourg Drop (LSD) model \cite{Ivanyuk2009} and $\delta W$ represents the effective shell-correction energy. The ground-state shell correction energies and the parametrization for $B_{\rm LSD}$ are taken from the mass table of Möller et al. \cite{Moller1995}. Furthermore, the dependence of the fission barrier on both the excitation energy $E^*_{\rm CN}$ and the angular momentum $\ell$ of the compound nucleus is taken into account as described in Ref. \cite{Giardina2018}, and the correction factor is expressed as
\begin{eqnarray}
f &=&h(T) \cdot q(\ell), \nonumber\\
h(T)&=&\{1+\exp[(T-T_0)/d]\}^{-1}, \\
q(\ell)&=&\{1+\exp[(\ell-\ell_{1/2})/\Delta\ell]\}^{-1}.\nonumber
\end{eqnarray}
Here, $T=\sqrt{E^*_{CN}/a}$ is nuclear temperature, $d=0.3$ MeV is the rate of washing out the shell corrections with the temperature, $T_0=1.16$ MeV is the value at which the damping factor $h(T)$ is reduced by $1/2$;  $\Delta\ell=3\hbar$ is the rate of washing out the shell corrections with the angular momentum, $\ell_{1/2}=20\hbar$ is the value at which the damping factor $q(\ell)$ is reduced by $1/2$. The level-density parameter $a$, as described by Nerlo-Pomorska et al. \cite{Pomorska2006}, was used and it is given by the following expression:
\begin{eqnarray}\label{lvldensity}
a=0.092A&+&0.036A^{2/3}\mathfrak{B}_s+0.275A^{1/3}\mathfrak{B}_k\nonumber\\
&&-0.00146\frac{Z^2}{A^{1/3}}\mathfrak{B}_c,     
\end{eqnarray}
where $\mathfrak{B}_s$ is the surface term, $\mathfrak{B}_k$ is the curvature term and $\mathfrak{B}_c$ is the Coulomb term for a deformed nucleus \cite{Hasse}. All models and parameters used in calculating the survival probability $W^x_{\rm sur}(E^*_{x-1},\ell)$, are based on our previous decay studies for various reactions \cite{Kayumov2022, Nasirov2023, Nasirov2024, Nasirov2024232TH}. 

\section{Results and discussions}

\begin{figure}[ ]
\includegraphics[width=0.50\textwidth]{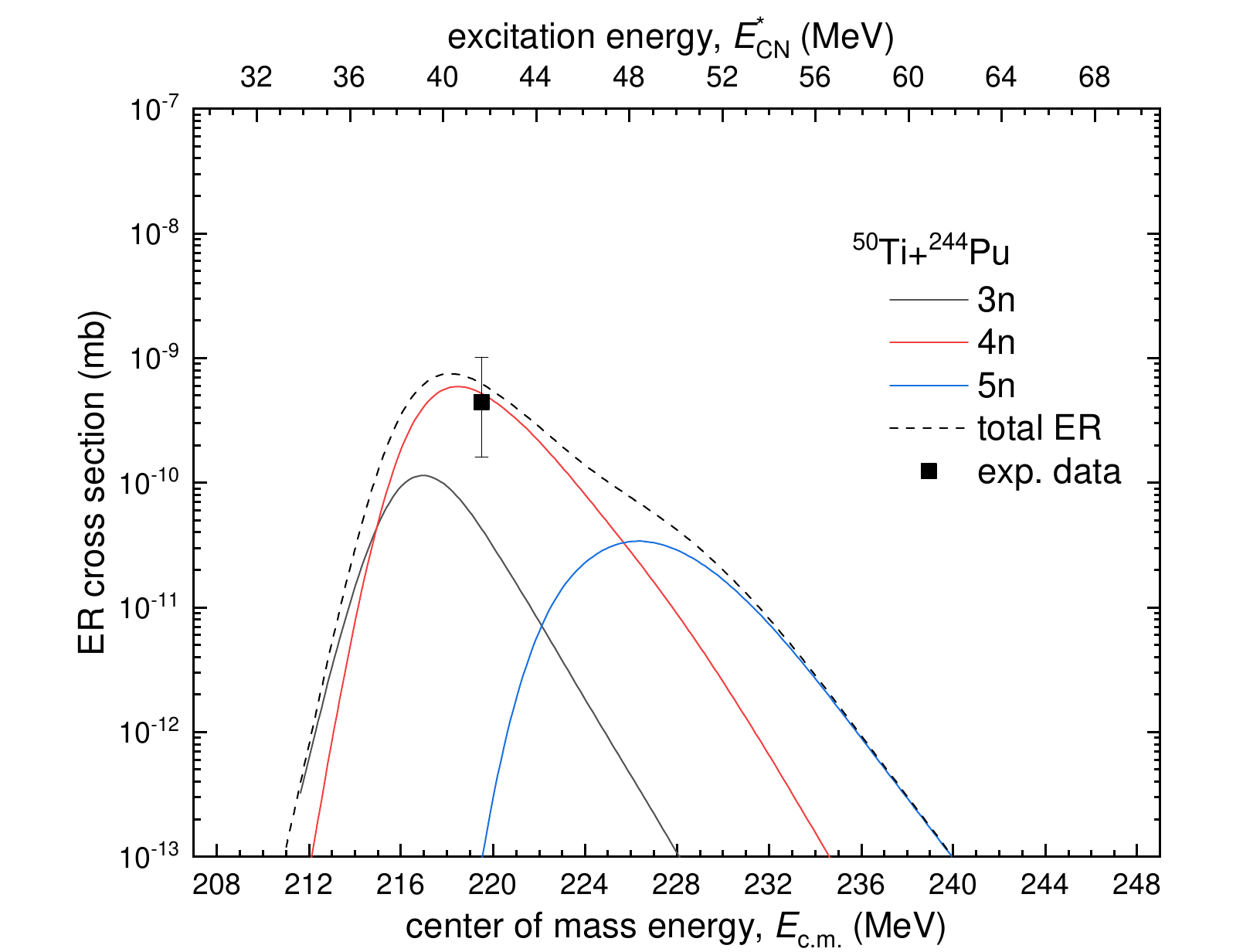}
\caption{(Color online) The calculated ER cross section for producing superheavy element 116 via $^{50}$Ti+$^{244}$Pu reaction in the 3n (solid black curve), 4n (solid red curve) and 5n (solid blue curve) channels. Dotted black line presents results for the total evaporation residue. The experimental data from in \cite{Gates2024}.}
\label{ERcross116}
\end{figure}

Recent experimental studies at Lawrence Berkeley National Laboratory’s 88-Inch Cyclotron focused on $^{244}$Pu($^{50}$Ti,$xn$)$^{294-x}$Lv reaction provides a measured production cross section of $0.44^{+0.58}_{-0.28}$ pb for the $4n$ channel at a center of the mass energy of 220(3) MeV \cite{Gates2024}. To provide theoretical insights into these experimental results, evaporation residue (ER) cross sections were calculated for the $^{50}$Ti+$^{244}$Pu reaction by using the dinuclear system (DNS) model. Figure \ref{ERcross116} presents the ER cross sections for $3n, 4n$ and $5n$ channels, as a function of both the center of the mass energy ($E_{\rm c.m.}$) and excitation energy ($E^*_{\rm CN}$). It can be seen from the results that the highest ER cross section is observed in the $4n$ channel with $\sigma_{ER}=0.58$ (pb), at the energy $E_{\rm c.m.} = 218.8$ MeV and the calculated results are in good agreement with the experimental data presented in \cite{Gates2024}. It can be seen from the figure that the optimal incident energy lies at $E_{c.m.}=218.1$ MeV. Presented theoretical results for different channels and OIE can be used as a prediction for future experiments on $^{50}$Ti+$^{244}$Pu reaction, and the maximum values for different channels are presented in Table \ref{tabERCS}. 

\begin{figure}[ ]
\includegraphics[width=0.50\textwidth]{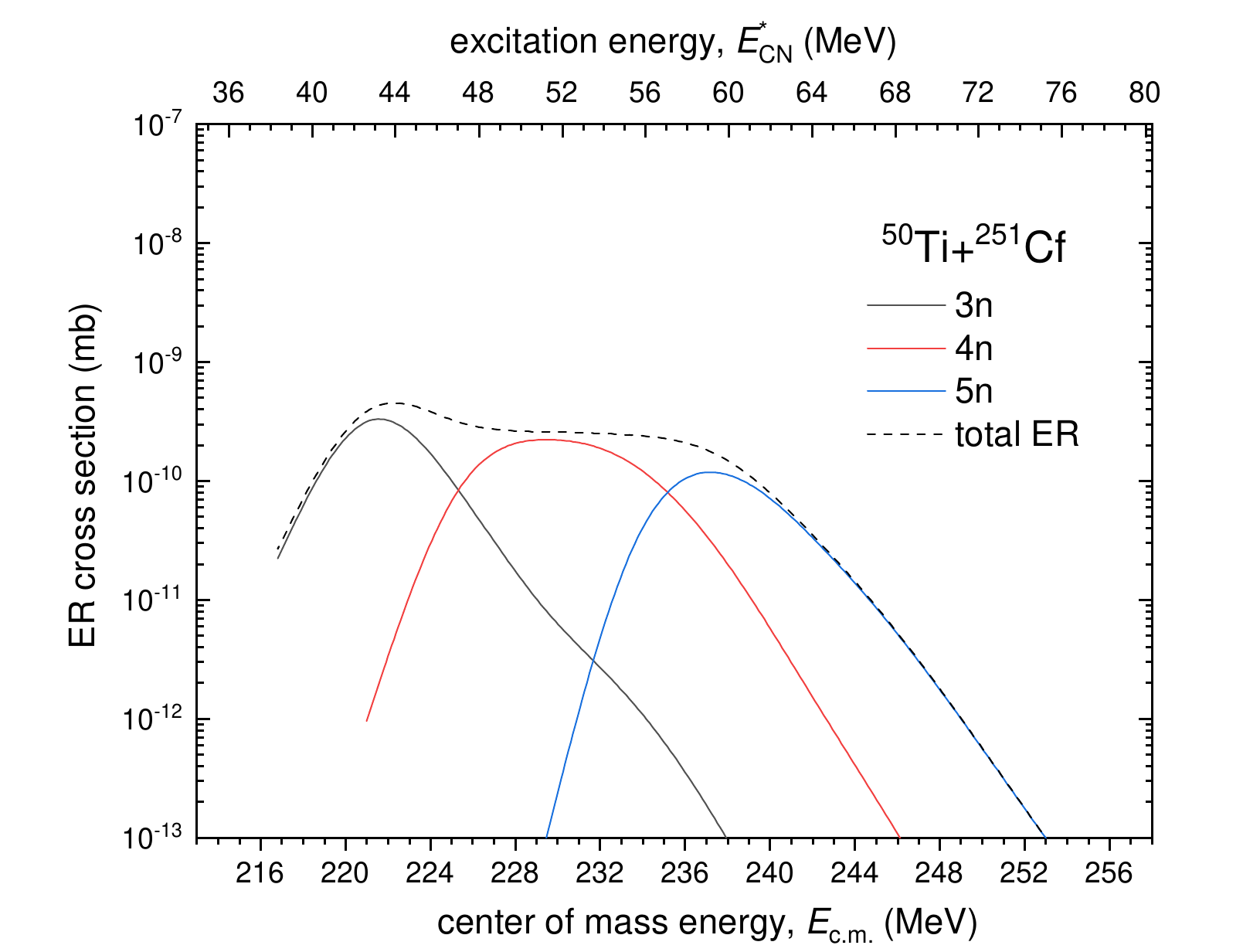}
\caption{(Color online) The same as figure \ref{ERcross116}, but $^{50}$Ti+$^{251}$Cf reaction where the 3n is solid black curve, 4n is solid red curve and 5n is solid blue curve. Dotted black line presents results for the total evaporation residue.}
\label{ERcross120}
\end{figure}

\begin{table*}[ ]
\caption{Comparison of the theoretical predicted ER cross sections with results presented in Refs. \cite{Gates2024} for the $^{50}$Ti+$^{244}$Pu reaction and predicted ER cross section for the $^{50}$Ti+$^{251}$Cf reaction.}
\label{tabERCS}
\begin{ruledtabular}

\begin{tabular}{cccccc}
Reaction &   & $E_{\rm c.m.}$ (MeV) &  $E_{\rm CN}^*$ (MeV)& This work $\sigma_{ER}$ (pb) &  \cite{Gates2024} \\ \hline
&  $3n$ & 216.5 &  38.3 & 0.13 & Exp. Results $4n$ \\
$^{50}$Ti$+^{244}$Pu&  $4n$ & 218.8 & 40.3 & 0.58 & at $E_{\rm CN}^*$=41 MeV  \\
& $5n$ & 226.4 &  48.2 & 0.034 &  $\sigma_{ER}$=0.44($^{+0.58}_{-0.28}$) pb \\
\hline
&  $3n$  & 221.6  &  43.4 & 0.34 &   \\
$^{50}$Ti$+^{251}$Cf &  $4n$  & 229.4  &  51.2 & 0.23 &  $-$ \\
& $5n$   & 237.1  &  58.9 & 0.12 &   \\
\end{tabular}
\end{ruledtabular}
\end{table*}

Figure \ref{ERcross120} shows results of calculated ER cross section for the $^{50}$Ti+$^{251}$Cf reaction as a function of energies. The results indicate distinct peak cross sections for the neutron evaporation channels (3n, 4n, and 5n), which are crucial for evaluating the probability of synthesizing element 120 (detailed results in Table \ref{tabERCS}). 
Previous theoretical studies \cite{Wang2012,Liu2016}, have estimated the ER cross sections for the $^{50}$Ti+$^{251}$Cf reaction be within the range of $10^{-10}-10^{-12}$ mb, depending on the neutron evaporation channel. Cross sections computed in this work fall within this range, reinforcing the consistency of DNS approach. However, the present study finds higher maximum ER cross section values, reaching 0.34 pb for the 3n channel and 0.23 pb for the 4n channel, as detailed in Table \ref{tabERCS}. These higher values suggest an increased probability of forming 120 nuclei compared to earlier calculations. Additionally, OIE for $^{50}$Ti+$^{251}$Cf reaction can be extracted from figure \ref{ERcross120} and its value equal to $E_{c.m.}=222.3$ MeV. 

Compared to previous studies \cite{Wang2012,Liu2016}, these calculations indicate a slight shift in optimal incident energy positions for the ER cross sections. The energy shift may be influenced by dynamical calculation of the energy ranges leading to the capture and deformations of the target $^{251}$Cf nucleus, which affects the compound nucleus formation probability and shifts the optimal energy window. Author \cite{Long022030} predicted optimal incident energies for $^{50}$Ti induced reaction, with target $^{249}$Bk and $^{249}$Cf, $E_{c.m.}=224.9$ MeV and $E_{c.m.}=229.8$ MeV, correspondingly.  

A key motivation for investigating the $^{50}$Ti+$^{251}$Cf reaction is to evaluate its possibility for synthesizing element 120. Previous studies \cite{Wang2012, Liu2016, Nasirov2011, Niu2021} have primarily focused on the $^{50}$Ti+$^{249}$Cf reaction to synthesis 120 element, and results presented in this work for $^{50}$Ti+$^{251}$Cf reaction may be more favorable for producing element 120. The present calculation shows greater ER cross section values compared to predictions for the $^{50}$Ti+$^{249}$Cf reaction. This suggests an increased probability of forming element 120 via neutron evaporation in the $^{50}$Ti+$^{251}$Cf system. This could be attributed to fusion barrier height which is used to calculate fusion probability, and the treatment of fission barrier modifications with excitation energy and orbital angular momentum on calculation of survival probability. For the reaction $^{50}$Ti+$^{249}$Cf, which is presented in \cite{Nasirov2011}, fusion probability is less than the results presented in this work, and this is because in $^{50}$Ti+$^{249}$Cf reaction has greater fusion barrier than $^{50}$Ti+$^{251}$Cf, which reduces the probability of formation compound nucleus.

\section{Conclusions}

The present study investigates the formation of superheavy elements with atomic numbers 116 (Livermorium) and 120 through the $^{50}$Ti+$^{244}$Pu and $^{50}$Ti+$^{251}$Cf reactions. By employing the dinuclear system model,  key reaction parameters such as: capture cross section, fusion probability, and evaporation residue cross sections, are systematically analyzed. The results provide crucial insights into the probability of synthesizing these elements and the factors influencing on their formation.

Calculations demonstrate that the $^{50}$Ti+$^{244}$Pu reaction exhibits a higher fusion probability than the $^{50}$Ti+$^{251}$Cf system, primarily due to the lower fusion barrier associated with the entrance channel. 

The theoretical $4n$ evaporation residue cross section for $^{50}$Ti+$^{244}$Pu ($\sigma_{ER} = 0.58$ pb at $E_{c.m.} = 218.8$ MeV) agrees well with recent experimental data, confirming the predictive power of DNS approach. The $^{50}$Ti+$^{251}$Cf reaction, while displaying a lower fusion probability, still shows significant ER cross-section, can be considered as an alternative reaction channel for synthesizing element 120. Results presented in this work provides OIE for $^{50}$Ti+$^{244}$Pu reaction at $E_{c.m.}=218.1$ MeV and for $^{50}$Ti+$^{251}$Cf reaction at $E_{c.m.}=222.3$ MeV.

Additionally, this study highlights the impact of nuclear structure effects, such as deformation parameters, on the synthesis of superheavy elements. The orientation dependent fusion probabilities indicate that nuclear shape and alignment, play a significant role in determining the overall reaction outcome. The findings emphasize the importance of selecting optimal beam energies and target-projectile combinations to maximize the synthesis probability of new elements in the island of stability.

The results presented in this work provide valuable theoretical guidance for future experimental efforts aimed at expanding the superheavy element landscape. 

\section{Acknowledgments}
Author acknowledges Nasirov A.K. for a useful discussion and the guidance through this work. 

\bibliography{references}

\end{document}